\documentclass[twocolumn,notitlepage,floats,nofootinbib,amsmath,amssymb,aps,prd]{revtex4-1}

\usepackage{amsmath,amssymb,amsfonts,latexsym,cancel,amsthm}

\usepackage{mathtools}
\usepackage{graphicx}
\usepackage{color}
\usepackage{amsbsy}
\usepackage{hyperref}
\usepackage{tikz}

\usepackage{seqsplit}
\usepackage{comment}

\newcommand{\be}{\begin{equation}}
\newcommand{\beq}{\begin{equation}}
\newcommand{\ee}{\end{equation}}

\def\bea {\begin{eqnarray}}
\def\eea {\end{eqnarray}}

\def\dd{{\rm d}}

\definecolor{newgreen}{rgb}{0.0, 0.75, 0.0}
\definecolor{cadmiumgreen}{rgb}{0.0, 0.42, 0.24}

\newtheorem*{theorem}{Theorem}

\begin{document}

\title{Gentle spaghettification in effective LQG dust collapse }

\author{Francesco Fazzini} \email{francesco.fazzini@unb.ca}
\affiliation{Department of Mathematics and Statistics, University of New Brunswick, \\
Fredericton, NB, Canada E3B 5A3}

\begin{abstract}

Effective dust collapse inspired by loop quantum gravity predicts two main features for general realistic initial profiles: a quantum gravitational bounce of the stellar core, when the energy density becomes planckian, and shell-crossing singularities, arising within almost a planckian time after the bounce. The aim of this work is to study the mathematical and physical features of the effective spacetime near these singularities, through the Jacobi deviation equation. The results show that the radial and angular deviations between dust particles remain finite at both the bounce and shell-crossing singularity, providing a physical justification for extending the spacetime beyond shell-crossing singularities through weak solutions.

\end{abstract}

\maketitle

\section{Introduction}

Spherically symmetric stellar collapse in classical general relativity generally leads to the formation of crush singularities, also called central singularities, according to the Penrose singularity theorem \cite{{Penrose:1964wq}}. Even though such singularities have been widely studied in the literature and their mathematical properties analyzed in great detail, the common belief is that they are a pathological feature of Einstein theory of gravity, rather than a physical prediction. It has also been found that such singularities are not the only ones that develop during gravitational spherically symmetric collapse for generic initial density profiles. If neighboring layers of the collapsing star move at significantly different speeds, large inhomogeneity gradients can develop in the stellar energy density profile, sometimes leading to the so-called shell-crossing singularities (SCS) \cite{Penrose:1999vj,Szekeres:1995gy,Hellaby:1985zz,Lasky:2006hq}. Although shell-crossing singularities are also physical (i.e., not mere coordinate artifacts), much less (though still significant) effort has been devoted to studying SCS compared to crush singularities, and the reasons are mainly threefold: firstly, during the collapse of pressureless matter, Hellaby and Lake proved \cite{Hellaby:1985zz} that one can always choose initial data that do not develop such singularities before the central singularity is reached, which, in turn, is generally unavoidable. Additionally, by examining the mathematical features of these singularities, one can show \cite{Szekeres:1995gy} that curvature scalars diverge less strongly compared to those associated with the crush singularities. Moreover, the behavior of matter falling toward these singularities is much less dramatic: despite the tidal forces being divergent at SCS, the angular and radial displacement of test particles remain finite \cite{Szekeres:1995gy}. Furthermore, since the radial tidal force is negatively divergent, the so-called spaghettification effect is absent near SCS.\par
The purpose of this work is to understand how this picture changes once quantum gravitational corrections are taken into account. Among the possible candidate theories of quantum gravity, loop quantum gravity is one of the best-developed approaches \cite{Rovelli:2004tv,Thiemann:2007pyv}. Although a quantum gravitational model of stellar collapse is not yet available from the full theory, effective models have developed in recent years, where quantum gravitational corrections to Einstein equations are inspired by features of the underlying fundamental theory. The effective model on which this work is based is inspired by effective loop quantum cosmology (LQC), where the classical big bang singularity is replaced by a quantum gravitational bounce \cite{Ashtekar:2006wn}, arising when the energy density becomes planckian.

Modifications to Einstein's equations at the effective level arise from a loop quantization of the midisuperspace, with spherical symmetry imposed at the classical level. In LQG-inspired models, such modifications typically affect only the geometrical part of the field equations, while the matter sector remains unaltered---a strategy that is also followed in the present work. This approach reflects a widely adopted methodology in Loop Quantum Cosmology (LQC), where the holonomy of the extrinsic curvature is quantized---referred to as $K$-loop quantization---in contrast to the full theory, where the holonomy of the entire Ashtekar-Barbero connection is quantized \cite{Bojowald:2003mc,Vandersloot:2006ws,Singh_2013}. In this treatment, inverse triad corrections are neglected. The physical length over which the holonomy is computed is determined via the improved dynamics scheme (the so-called $\Bar{\mu}$-scheme), which ensures that quantum corrections are suppressed in the low-curvature regime \cite{Ashtekar:2006wn} and connects the effective theory to the minimum non-zero area eigenvalue of the full quantum theory $\Delta \propto l_{Pl}^2$. 

This quantization prescription has been implemented explicitly in \cite{Husain:2022gwp}, where both the areal gauge and dust-time gauge are fixed at the classical level. However, the dynamics is analyzed only at the effective level. The expectation that these effective equations reliably capture the underlying quantum dynamics stems from an argument inherited from LQC. Specifically, it has been shown \cite{Rovelli:2013zaa} that for cosmological solutions with physical volumes at the bounce satisfying $V_{phys}\gg \Delta^{\frac{3}{2}}$
 , quantum fluctuations in the gravitational field become negligible even when the curvature approaches the Planck scale. This explains why quantum and effective cosmological dynamics agree so well numerically when large physical volumes are considered.

To apply the same reasoning to gravitational collapse, consider the simplest case: Oppenheimer–Snyder (OS) collapse. The interior solution of the OS model is isomorphic to a cosmological spacetime, and the physical volume of the star at the bounce is much larger than Planckian \cite{Husain:2021ojz}. Therefore, quantum fluctuations are also expected to be negligible in this astrophysical context. It is important to note, however, that this argument may not hold in the case of inhomogeneous collapse, where large inhomogeneities in the energy density develop---the scenario explored in this work. A comment on this limitation is provided in the concluding discussion. Motivated by feasibility and also by the previous argument, the effective dynamics has been extensively studied in recent years, both in the dust \cite{Husain:2021ojz,Han:2023wxg,Lewandowski:2022zce,Giesel:2023hys,Fazzini:2023scu,Husain:2022gwp,Kelly:2020lec,Munch:2020czs,Munch:2021oqn,Cafaro:2024vrw,Fazzini:2023ova,Cipriani:2024nhx,Bobula:2023kbo,Bobula:2024chr,Giesel:2023tsj} and pressure cases \cite{Wilson-Ewing:2024uad,Cafaro:2024lre}. For a comparison with other existing effective models, see \cite{Ashtekar:2023cod}.

Despite much effort being devoted within this scheme to the Oppenheimer-Snyder collapse, where the stellar dynamics follows exactly the cosmological one \cite{Munch:2020czs,Munch:2021oqn,Lewandowski:2022zce, Fazzini:2023scu,Giesel:2023hys}
the general picture regarding initial continuous decreasing inhomogeneous profiles is different \cite{Fazzini:2023ova}: the collapsing stellar core undergoes a quantum gravitational bounce when its energy density becomes planckian. Immediately after that, shell-crossing singularities arise (which, as mentioned before, are physical singularities), and the equations of motion break down. Shell-crossing singularities produce a non-isolated thin shell of matter, which dynamics can in principle be studied with a dynamical version of the Israel junction conditions \cite{Fazzini:2025zrq}, a tool that is not yet available.

The question this work aims to address is how tidal forces behave during this effective dynamics, focusing in particular on the bounce point and shell-crossing singularities. The main results we found are the following: the angular tidal forces remain bounded during the bounce of the stellar core and reach their maximum exactly at the bounce point. Unlike the classical case, there is a short period of time around the bounce point, in which the angular tidal force is positive, and two zeroes are attained, just before and after the bounce. The radial component of the tidal force, instead, is also bounded at the bouncing point but
diverges at SCS. However, unlike the classical case, where the tidal forces negatively diverge at SCS, in the effective case, the radial component of the tidal force is positive at SCS, meaning that spaghettification of a test body occurs, similarly to what happens near the classical central singularity. Despite this, importantly, by solving the radial component of the geodesic deviation equation, one finds that the radial separation between two points remains finite at the shell-crossing singularity, meaning that the spaghettification does not result in an infinite stretching of the body. This allows us to conclude that such singularities, arising during effective stellar collapse, should not be considered as dramatic as the central classical singularities, and it motivates the extension of spacetime beyond these physical singularities.

\section{Dust collapse in the marginally bound case}

Effective dust collapse within the $\Bar{\mu}+K-$loop quantization scheme has been extensively studied in Painlevè-Gullstrand coordinates \cite{Husain:2021ojz,Kelly:2020lec,Husain:2022gwp,Fazzini:2023scu,Cafaro:2024vrw,Cipriani:2024nhx,Fazzini:2025hsf} and Lemaître-Tolman-Bondi coordinates (LTB) \cite{Giesel:2023hys,Fazzini:2023ova,Bobula:2024chr}.

Although the original formulation of the model---both at the quantum and effective levels---was carried out within the dust-time and areal gauges, it has been shown that the areal gauge condition can be relaxed at the classical level, and imposed directly at the effective level. This leads to an effective model featuring an unmodified classical diffeomorphism constraint, in line with the structure of the full theory \cite{Giesel:2023hys}. This relaxation enables the study of dynamics in various spatial gauges at the effective level, with the effective equations of \cite{Husain:2022gwp} recovered upon imposing the areal gauge. The flexibility in choosing the spatial gauge at the effective level is particularly valuable when analyzing the formation and physical properties of shell-crossing singularities \cite{Fazzini:2023ova}. For this reason, the results presented in this work make use of LTB coordinates.

The line element describing effective stellar collapse in these coordinates takes the following form
\begin{equation}
 \dd s^2=-\dd t^2 + \frac{(\partial_R r(R,t))^2}{1+\varepsilon(R)}\dd R^2 +r(R,t)^2 \dd \Omega^2 ~,
 \label{metric}
\end{equation}
where $\varepsilon$ is a time-independent function fixed by the initial condition, and determines the Newtonian energy (kinetic plus potential) of any shell $R$ at infinity. We will focus here on the marginally bound case $\varepsilon(R)=0$, $\forall R$. The function $r(R,t)$ is the areal radius of each shell $R$ of the distribution at time $t$, and its explicit form is provided by the solution of the effective equations written in the LTB gauge.
In turn, the equation of motion in LTB coordinates takes the following form \cite{Giesel:2023hys}:
\begin{equation}
    \left( \frac{\Dot{r}}{r}\right)^2=\frac{F(R)}{r^3} \left(1-\frac{F(R)\gamma^2 \Delta }{r^3} \right) ~,
    \label{beautiful}
\end{equation}
where we defined $F(R)\equiv 2Gm(R)$ for compactness, and
 $m(R)$ is the ADM mass function for the portion of the star within the $R$-shell. The factor $\gamma$ is the Barbero-Immirzi parameter, that for simplicity we set to $1$. As one can explicitly notice from \eqref{beautiful}, the spacetime evolution is provided by the evolution of the areal radius of the shells composing the collapsing star.
The analytic solution of the previous equation reads:
\begin{equation}
r(R,t)=\left[F(R)\right]^{\frac{1}{3}}\left\{\frac{9}{4}\left[t-\alpha(R)\right]^{2} +  \Delta\right\}^{\frac{1}{3}}  ~,
\label{beauty}
\end{equation}
where $\alpha(R)$ is fixed by the initial condition. A fundamental feature of this generic solution, valid for any initial energy density profile, is that each shell $R$ will bounce at $t_{B}(R)=\alpha(R)$. In the pre-bounce (post-bounce) phase, we have $t<\alpha(R)$ ($t>\alpha(R)$). Even though from \eqref{beautiful} it seems that the shells' dynamics is decoupled, and each shell evolves independently from the others (influenced only by the gravitational mass within its spherical volume), such solutions generally lead to shell-crossing singularities, at which point the equations cannot be trusted anymore. In order to define mathematically such singularities, we recall that the dust energy density in LTB coordinates reads
\cite{Giesel:2023hys}
\begin{equation}
\rho= \frac{\mathcal{H}_{dust}}{4 \pi\sqrt{q}} =\frac{\partial_R F}{ 8 \pi G r^2 \partial_R r}~. \label{rho}
\end{equation}

Here, $\sqrt{q}$ denotes the determinant of the spatial metric \eqref{metric}, and $\mathcal{H}_{dust}$ is the dust Hamiltonian density, integrated over $\dd \theta$ and $\dd \phi$, while $\rho$ represents the dust energy density. Note that these expressions retain their classical functional form, since the matter sector is kept classical in this effective model \cite{Kelly:2020lec}. As in the classical case, the LTB coordinates are adapted to the comoving dust frame. Therefore, $\rho$ corresponds to the $tt$ component of the energy-momentum tensor. Note that the functional relation between the lagrangian density for dust and the energy-momentum tensor is classical. Although the functional form of \eqref{rho} remains classical, its explicit behavior deviates from the classical solution when $\rho$ approaches the critical density $\rho_{crit.} = 3/(8 \pi G \Delta) \sim \rho_{\text{Pl.}}$, or equivalently when $r(R,t) \sim [F(R)\Delta]^{1/3}$---that is, near the bounce.

A dynamical solution of the form \eqref{beauty} develops a SCS if, for a given shell labeled by $R$, the following conditions are satisfied: $\partial_R F \neq 0$ and $\partial_R r(R,t) = 0$. From equation \eqref{rho}, it is easy to see that these conditions lead to a divergence in the dust energy density, which in turn results in a divergence of the curvature scalars \cite{Giesel:2024mps}.

\section{Choice of initial profiles}
The effective equations \eqref{beautiful} hold for any initial energy density dust configuration in the marginally bound case. In particular, one can easily derive the effective cosmological dynamics \cite{Ashtekar:2006wn} by requiring $\rho(t_0,R)=const.$, as well as the vacuum solution \cite{Giesel:2023hys,Giesel:2024mps}, by imposing $F=const$. Here, we are interested in stellar collapse, so we will assume that the initial energy density profile is positive and has compact support. Moreover, to keep the result as general as possible, we will consider decreasing inhomogeneous profiles ($\partial_R \rho(t_0, R)<0$) of compact support, without specifying the initial data.

For these kinds of profile, we can combine a theorem, stated and proved in \cite{Fazzini:2023ova}, and two results, one from \cite{Fazzini:2023ova} and the other from \cite{Fazzini:2025hsf}:
\begin{theorem}
    For the case $\varepsilon=0$, a shell-crossing singularity
forms if the initial distribution of the dust energy
density $\rho(R,t_0)$ is non-negative, continuous, of compact
support, and for which m(R) is not everywhere zero,
\end{theorem}
and:
\begin{enumerate}
    \item the latest time at which a shell-crossing singularity can arise in effective stellar collapse is $t=t_{B}(R)+(2/3)\sqrt{\Delta}$ \cite{Fazzini:2023ova}.
    \item for initially decreasing continuous profiles, shell-crossing singularities cannot arise in a time $t\leq t_{B}(R)$ \cite{Fazzini:2025hsf}.
\end{enumerate}
 Therefore, if we consider an initial profile that is non-negative, decreasing, and of compact support, a shell-crossing singularity will necessarily form for a certain shell $R$ at time $t_{B}<t\leq t_{B}+(2\sqrt{\Delta}/3)$. These results allow us to restrict the analysis of tidal forces to the post-bounce dynamics of the shells, particularly to almost a planckian time interval after the bounce.

\section{Geodesic deviation equation for LTB collapse}

In order to understand the tidal effects during generic dust collapse, we need to study the geodesic deviation equation. In the classical theory, given $u^{\mu}$ the 4-vector tangent to the geodesics, and $\delta x^{\nu}$, the displacement vector orthogonal to $u$ ($\delta x^{\nu}u_{\nu}=0$), the Jacobi equation holds \cite{Wald:1984rg}
\begin{equation}
\frac{D^2 \delta x^{\mu}}{D \tau^2}=R^{\mu}_{\nu \rho \sigma} u^{\nu} u^{\rho} \delta x^{\sigma} ~,
\label{jacobi0}
\end{equation}
up to linear order in the deviation and its derivative. The covariant derivative is computed along the direction individuated by $u^{\mu}$, and therefore $\tau$ is the proper time along the geodesics. This equation does not use the field equations (classical or modified), but comes from the geometrical definition of the Riemann tensor. Therefore, it also holds at the effective level.

We are interested in the evolution of the deviation between dust particles during the stellar dynamics, and since dust particles follow geodesics also at the effective level (which can be readily verified by checking that the dust four-velocity $u^\mu = (1, 0, 0, 0)$ satisfies the geodesic equation), \eqref{jacobi0} describes tidal forces experienced by the dust itself. 
In particular, we are looking at a congruence of time-like geodesics for initially free-falling dust particles. Since the fluid $4$-velocity coincides with the $4$-velocity of a radially free-falling test particle, equation \eqref{jacobi0} simplifies considerably
\begin{equation}
\frac{D^2 \delta x^{\mu}}{D \tau^2}=R^{\mu}_{0 0 \sigma} \delta x^{\sigma} ~.
\label{jacobi}
\end{equation}

In order to get rid of the covariant derivative and further simplify the equation, we use a common procedure and introduce tetrads \cite{dInverno:1992gxs}. The tetrads associated with the LTB line element \eqref{metric}, which coincide with the tetrad basis in the free-falling frame, in the marginally bound case $\varepsilon(R)=0$ take the following (classical) form:
\begin{align}
& e^{\mu}_0=\left\{1,0,0,0 \right\}   \\
&e^{\mu}_1=\left\{0,({\partial_R r})^{-1},0,0 \right\}\\
&e^{\mu}_2=\left\{0,0,r^{-1},0 \right\}\\
&e^{\mu}_3=\left\{0,0,0,({r\sin \theta})^{-1} \right\} ~.
\end{align}
These tetrads can easily be derived from $g^{\mu \nu}=e^{\mu}_{i} e^{\nu}_{j}\eta^{ij} $. Then, we write all the tensors appearing in \eqref{jacobi} in terms of tetrads:
\begin{align}
  &R^{\mu}_{\nu \rho \sigma}=R^{a}_{bcd}e^{\mu}_{a}e^{b}_{\nu} e^{c}_{\rho} e^{d}_{\sigma} ~, \\
&\delta x^{\mu}=e^{\mu}_{a}\delta x^{a} ~,
\end{align}
where $a,b,c,d$ run from $0$ to $3$. Notice that $\delta x^0=0$, since $u^\mu\delta x_{\mu}=0$ and $u^{\mu}$ have no spatial components. The deviation equation becomes
\begin{equation}
\frac{D^2 \delta x^{a}}{D \tau^2}=R^{a}_{0 0 b} \delta x^{b} ~,
\label{intermediate}
\end{equation}
which $0-$component is trivially satisfied. To arrive at \eqref{intermediate} we used the tetrad postulate, which can be written as $De^{a}_{\mu}/D\tau=0$.
The great advantage of using tetrads in LTB coordinates is that they define a locally inertial freely falling frame, for which the covariant derivative reduces to a simple ordinary derivative \cite{Szekeres:1995gy,Fuchs(1990)}. This leads to
\begin{equation}
\frac{\dd^2 \delta x ^{a}}{\dd \tau^2}=R^{a}_{00 b } \delta x^{b}    ~.
\end{equation}
To conclude, since the LTB coordinate time is the proper time of free-falling timelike observers, we can set
$\tau=t$, and the equation further simplifies:
\begin{equation}
\frac{\partial^2 \delta x ^{a}}{\partial t^2}=R^{a}_{00 b } \delta x^{b}    ~.
\label{jacobi1}
\end{equation}
Although the explicit form of the effective Riemann tensor differs from the classical one, the effective LTB metric does not acquire quantum correction in its functional dependence from the solution $r(R,t)$, and the same holds for the Riemann tensor. Therefore, we can use the classical result obtained in \cite{Joshi:2024djy} to write down \eqref{jacobi1} in a more explicit form:
\begin{align}
 &\partial^2_t{ \delta x ^1}=\frac{\partial^2_t \partial_R{r}}{\partial_R r}\delta x^{1} ~,\label{1}\\
& {\partial^2_t \delta x ^{i}}=\frac{\partial^2_t{r}}{ r}\delta x^{i} ~, \label{2}
\end{align}
where the first equation determines the radial deviation
and the index $i=2,3$ indicates the angular deviations in the tetradic basis. The time component of \eqref{jacobi1} gives a trivial identity. It is worth mentioning that even though the functional dependence on $r$ takes the classical form, since the effective solution differs from the classical one, the same clearly holds for the solutions of equations \eqref{1} and \eqref{2}. 

\section{Tidal forces during effective stellar collapse}

As pointed out before, effective inhomogeneous dust collapse does not end in the classical central singularity, which is replaced by a quantum gravitational bounce. In classical Einstein theory, the angular tidal forces close to shell-crossing singularities do not diverge, while despite radial tidal forces being divergent, the radial deviation does not go to zero at SCS, nor does it diverge \cite{Szekeres:1995gy}.
This makes such singularities less pathological than the central one. Here, we show that similar results also hold at the effective level.

To prove this, we compute explicitly \eqref{1}, \eqref{2}, using the solution \eqref{beauty} of the effective equation \eqref{beautiful}. Let us start from the equation for the angular deviation. After some computation, we find
\begin{equation}
\partial^2_t \delta x^{i}=\frac{3\left[\Delta-\frac{3}{4}(t-\alpha)^2 \right]  }{2 \left[\frac{9}{4}(t-\alpha)^2+\Delta \right]^2  } \delta x^i ~.
\label{attained}
\end{equation}

Firstly, we note that in the classical case $\Delta \rightarrow 0$, the angular tidal forces negatively diverge for $t\rightarrow \alpha(R)$ (classical central singularity). In contrast, here it remains always bounded, reaches its maximum at $t=\alpha(R)$ (the bouncing point) and has two zeros, for $t=\alpha \pm 2\sqrt{\Delta}/\sqrt{3}$.

The departure from the classical focusing behavior is clearly due to quantum gravity repulsion, which weakens the angular tidal force and, for almost a planckian time, produces a tidal angular repulsion.
\begin{figure}
    \centering
    \includegraphics[scale=0.6]{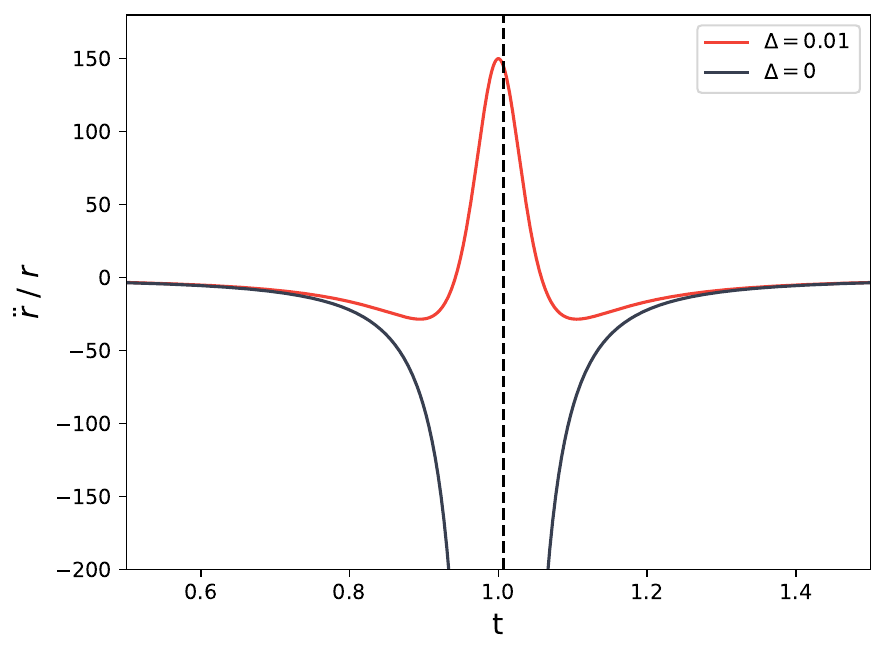}
      \caption[]{\footnotesize
 Angular tidal force for $\alpha=1$, in the classical (dark blue curve) and effective collapse (red curve). The dotted vertical line signals the latest time at which a SCS occurs in the effective case.} In the classical case the function diverges at the crush singularity $t=\alpha=1$, while remains bounded at the effective level.  \label{figure}
\end{figure}
Fig.\ref{figure} shows the comparison between the classical and effective cases. Notice that angular tidal forces do not diverge at shell-crossing singularity. Although the latest time at which a SCS arises and the largest zero of \eqref{attained} are quantitatively similar, they must be considered as two distinct features.

 It is worth mentioning that even though the angular tidal force remains bounded, it reaches a planckian value at the bounce, given by
 \begin{equation}
\partial^2_t \delta x^i |_{bounce}  =\frac{3}{2\Delta} \delta x^i   ~.
 \end{equation} 
 
 Let us now examine the radial tidal force. By explicitly writing the right-hand side of equation \eqref{1} (divided by $\delta x^1$), after a lengthy but straightforward computation, we find (see Appendix \ref{AppendixB})
\begin{equation}
\frac{\partial^2_t \partial_R r}{\partial_R r}= \left(\frac{F}{r^3}-\frac{10 \Delta F^2}{r^6} \right)+\frac{\partial_R F}{\partial_R r}\left( \frac{4\Delta F}{r^5} -\frac{1}{2r^2} \right)~. 
\end{equation}
Firstly, we notice that divergence can only occur if $\partial_R r=0$, provided that the numerator of the last term is non-zero. This indicates that the radial tidal force remains finite at the bounce, while it instead diverges at SCS.
To study the radial tidal force near and at SCS, we can assume that the first term is negligible
\begin{equation}
\frac{\partial^2_t \partial_R r}{\partial_R r}\sim \frac{\partial_R F}{2 r^5\partial_R r}\left( 4\Delta F -r^3 \right)~.
\label{tidalapprox}
\end{equation}

Next, we recall that for the profiles we are considering (decreasing, continuous, and of compact support), shell-crossing singularities arise for $ \alpha< t \leq \alpha + 2\sqrt{\Delta}/3$, which implies $F\Delta <r^3\leq 2F\Delta$, from \eqref{beauty}. This means that the numerator of \eqref{tidalapprox} is strictly positive and that at the shell-crossing singularity the radial tidal force diverges. It is worth noting that such a divergence arises in the deep quantum gravitational regime and is a positive divergence: an infinite stretching radial force arises, conceptually similar to the one occurring classically near the central singularity, producing the \emph{spaghettification effect}. In contrast, near classical shell-crossing singularities, the radial tidal force is negatively divergent: two points radially separated tend to be crushed together by the tidal force, even though in a weak fashion \cite{Szekeres:1995gy}, as one can see by setting $\Delta = 0$ in \eqref{tidalapprox}.

Let us now focus on the (positive) divergent behavior of the radial tidal force near and at the SCS. The condition for SCS can be derived from the solution \eqref{beauty} and rewritten as \cite{Fazzini:2023ova}
\begin{equation}
(t-\alpha)^2\partial_R m-2m(t-\alpha)\partial_R \alpha+\frac{4 \Delta}{9}\partial_R m=0 ~,
\label{doble}
\end{equation}
see also Appendix \ref{AppendixC}.
This is a second-degree equation in $t-\alpha$, with solutions $t_{1}=\alpha(R)+\delta t_1(R)$, $t_{2}=\alpha(R)+\delta t_2(R)$, with $\delta t_1 \leq \delta t_2$.
These solutions are guaranteed to be real and positive for continuous, inhomogeneous, decreasing profiles (for which $\partial_R \alpha >0$ \cite{Fazzini:2025hsf}) of compact support, as a consequence of the results in the previous section. It can also be easily shown that the solutions are distinct \cite{Fazzini:2023ova}, except for an initial condition that satisfies $\frac{m\partial_R \alpha}{\partial_R m}=\frac{2\sqrt{\Delta}}{3}$, which is a fine-tuned case which we do not consider here.
Therefore, we can rewrite the condition for SCS as
\begin{equation}
 (t-\alpha(R)- \delta t_1 (R))(t-\alpha(R)-\delta t_2(R))=0 ~.  
\end{equation}
This means that close to $t=t_{1}=\alpha + \delta t_1$, the function $r'$ goes to zero linearly: $\partial_R r\sim K(t_{1},t_2,R) (t_{1}-t)$ for $t\lesssim t_{1}$, with $K(t_{1},t_2,R)>0$ (see Appendix \ref{AppendixC}). Here we are only interested in the tidal forces before and at the shell-crossing singularity. Beyond this point, the analytic solution \eqref{beauty} is not reliable. With this in hand, we can rewrite \eqref{1}:
\begin{equation}
\partial^2_t \delta x^1 \sim \frac{B(t_1,t_2,R)}{t_{1}-t}\delta x^1 ~,  
\label{approx3}
\end{equation}
where $B(t_{1},t_2,R)$ is given by $\partial^2_t \partial_R r$ evaluated at shell-crossing singularity divided by $K(t_1,t_2,R)$. It is important to keep in mind that this approximation holds only close to the SCS ($t\lesssim \alpha(R)+\delta t_1(R)$). Equation \eqref{approx3} can be solved analytically. Let us call $t_1=t_{SCS}$, and define $\delta t\equiv t_{SCS}-t>0$. The solution is
 \begin{align}
  \delta x^1 (\delta{t})\sim & c_1(R) \sqrt{\delta{t}}  ~I_1\left[2 \sqrt{B(t_{SCS},R)\delta{t}}\right]+ \\
 &c_2(R) \sqrt{\delta{t}} ~K_1\left[2\sqrt{B(t_{SCS},R)\delta{t}}\right]  ~,
 \end{align}
 where $I_1$ and $K_1$ are modified Bessel functions of the first and second kinds, and $c_1(R),c_2(R)$ are time independent functions determined by initial conditions. The previous solution holds only for $\delta{t}\rightarrow 0^+$. While the first term goes to zero in this limit, the second term approaches a constant  non-zero value. Their explicit Taylor expansions are:
\begin{align}
\sqrt{\delta{t}}I_1\left[2 \sqrt{B(t_{SCS},R)\delta{t}}\right] =  \sqrt{B(t_{SCS},R)}\delta{t}
 + O(\delta{t}^2)~, \\
 \sqrt{\delta{t}}K_1\left[2 \sqrt{B(R)\delta{t}}\right] = \frac{1}{2\sqrt{B(t_{SCS},R)}}+O(\delta{t})~. 
 \end{align}
Therefore, even though the radial tidal forces diverge at SCS, two neighboring points that are radially separated (and thus belong to two different matter layers) do not generally clash when a shell-crossing singularity occurs. For example, these points can be thought of as belonging to two shells close to the ones that cross, but with a larger coordinate separation $\delta R$). If these points belong exactly to the crossing shells, they will be crushed, and this condition is given by $c_2(R)=0$ \cite{Szekeres:1995gy}, since at SCS $\partial_R r(R,t_{SCS})=0$, that implies $\delta r=\delta x^1=0$.
This implies that the radial portions of the collapsing star (or even geodesic test particles radially separated) undergo spaghettification, but without being infinitely stretched. We refer to this effect as \emph{gentle spaghettification}. 

This result leads to the conclusion that SCSs, which generally arise in effective stellar collapse, are less pathological than classical crush singularities. It also justifies the extension of the spacetime beyond SCS \cite{Husain:2022gwp,Cipriani:2024nhx,Fazzini:2025hsf,Liu:2025fil,Fazzini:2025zrq,Husain:2025wrh}. It is important to emphasize that this result is nontrivial. As mentioned above, even though the structure of the deviation equation follows the classical form, the effective solution differs from the classical one. Therefore, the behavior of tidal forces at shell-crossing singularities also differs from the classical behavior in the effective model.

To conclude, for profiles that develop shell-crossing singularities in the pre-bounce phase (where the classical Einstein equations hold), the classical analysis \cite{Szekeres:1995gy} can be directly exported to the effective model.

\section{conclusions}

Shell-crossing singularities are physical singularities generally unavoidable in effective dust collapse inspired by loop quantum gravity, within the $\Bar{\mu}+K$-loop quantization scheme. They therefore play a central role in this effective model, and their geometrical and physical features need to be investigated. In this work, we focus on the analysis of the tidal forces arising from effective dust collapse, with particular attention to the bouncing point and shell-crossing singularities. We found that tidal forces remain bounded at the bounce, in contrast to what happens in the classical case at the central crush singularity, and the same holds for the angular component at the shell-crossing singularity. However, the radial component of the tidal force positively diverges at the shell-crossing singularity.

Nevertheless, when studying the behavior of the radial deviation by solving the Jacobi equation, we find that it remains bounded throughout the dynamics, also when the shell-crossing singularity arises. The positive divergent behavior of the tidal force implies a spaghettification effect, similar to the one occurring near the central singularity in classical stellar collapse. However, the bounded behavior of the radial displacement between nearby points makes such an effect much less dramatic. We refer to this phenomenon as \emph{gentle spaghettification}, and it allows the dynamics to extend beyond shell-crossing singularities in a meaningful way, through the integral form of the equations of motion. Future work in this direction will aim to investigate the tidal forces in the effective gravitational collapse of dust in the non-marginally bound case \cite{Cipriani:2024nhx}, and of fluids with pressure \cite{Cafaro:2024lre}, in order to understand, at least at the numerical level, the role of pressure in the tidal forces experienced by test particles.

Although this work does not aim to study the spacetime evolution beyond the formation of shell-crossing singularities, the analysis presented here could be applied to spacetimes where such an extension has been performed, particularly to investigate tidal forces near the non-isolated thin shell generated by the SCS. When SCSs develop, the effective equations expressed in Painlevé-Gullstrand (PG) coordinates exhibit characteristic crossing, which imply a multi-valued behavior of the extrinsic curvature. Since these equations take the form of hyperbolic conservation laws in the marginally bound case \cite{Husain:2022gwp} and balance laws in the non-marginally bound case \cite{Cipriani:2024nhx}, one might be tempted to study their dynamics through weak solutions, as commonly done in other fields where characteristic crossings occur \cite{Batchelor,Hoefer2006}. This approach results in an evolving shock in the extrinsic curvature \cite{Husain:2022gwp,Cipriani:2024nhx}.

However, it was recently shown \cite{Fazzini:2025zrq} that the integral formulation underlying weak solutions implicitly assumes the continuity of PG time throughout the shock (the SCS in LTB coordinates), which is physically questionable. Indeed, the Misner-Sharp mass function becomes discontinuous at the shock, implying different time dilations measured by interior and exterior metrics, and consequently different PG times flowing. Furthermore, weak solutions exhibit superluminal behavior. The proper way to extend the spacetime beyond the SCS is to apply the Israel junction conditions to a non-isolated thin shell, ensuring subluminal propagation and allowing PG time to be discontinuous at the shock. Once this framework is established, understanding the evolution of tidal forces near the evolving thin shell will be crucial.

To conclude, a natural question arises as to whether shell-crossing singularities can be avoided once a full quantum theory of dust collapse is taken into account, going beyond the effective approach. In principle, quantum gravitational fluctuations in the fields could introduce corrections to the effective equation \eqref{beautiful} that prevent the formation of SCS and lead to a singularity-free evolution, although this outcome is not guaranteed. If this scenario holds, the thin shell dynamics beyond SCS studied here might qualitatively capture the behavior predicted by a more refined effective model. Moreover, since quantum fluctuations are expected to decay rapidly near the SCS, the tidal forces computed in this work could provide a reasonable approximation to those in such a hypothetical effective model.

\acknowledgments

This work is supported in part by the Natural Sciences and Engineering Research Council of Canada.

\appendix

\section{The radial tidal force}
\label{AppendixB}

We can compute the quantity $\frac{\partial_t^2 \partial_R r}{\partial_R r}$ from the general solution \eqref{beauty}. By deriving twice this expression in time, and rewriting the result in term of $r(R,t)$, we get
\begin{equation}
 \partial_t^2 r(R,t)= \frac{F(R)^{\frac{2}{3}}}{r^4}\left[r^3+2\Delta F(R) \right] ~.
\end{equation}
Then, deriving with respect of the radial coordinate, one gets
\begin{equation}
\partial_R{\partial_t^2 r(R,t)}= \frac{2\Delta}{r^4}\partial_R F^{\frac{5}{3}}+\frac{\partial_R F^{\frac{2}{3}}}{r}- F^{\frac{2}{3}}\left(1+\frac{8\Delta F}{r^3} \right)\frac{\partial_R r}{r^2} ~. 
\end{equation}
By dividing this expression by $\partial_R r(R,t)$, and rearranging terms, we get
\begin{equation}
\frac{\partial^2_t \partial_R r}{\partial_R r}= \left(\frac{F}{r^3}-\frac{10 \Delta F^2}{r^6} \right)+\frac{\partial_R F}{\partial_R r}\left( \frac{4\Delta F}{r^5} -\frac{1}{2r^2} \right)~. 
\end{equation}

\section{Behavior of $\partial_R r(R,t)$ close to SCS }
\label{AppendixC}

In this appendix, we prove that, close to SCS $t\lesssim t_1$, we can write $\partial_R r(R,t)=K(t_1,t_2,R)(t_1-t)+O(t_1-t)$, where $t_1(R)$ and $t_2(R)$ are the two real roots of \eqref{doble}, and $K(t_1,t_2,R)$ is a positive function. We can start by differentiating \eqref{beauty} with respect to $R$, getting
\begin{equation}
\frac{\partial_R r}{r}=\frac{\partial_R m}{3m}+\frac{3}{2}\frac{(\alpha(R)-t)\partial_R \alpha}{\left[\frac{9}{4}(\alpha-t)^2+\Delta \right]}  ~,
\end{equation}
which can be rewritten as 
\begin{equation}
\partial_r(R,t)=\frac{3r}{4m}\frac{\partial_R m (\alpha-t)^2+\frac{4}{9}\Delta \partial_R m-2\partial_R \alpha(t-\alpha)}{\frac{9}{4}(t-\alpha)^2+\Delta} ~.
\end{equation}
We recognize the last term in the numerator as the left-hand side of \eqref{doble}, and therefore we can rewrite the previous equation as
\begin{equation}
    \partial_R r(R,t)=\frac{3r}{4m}\frac{[t-t_1(R)][t-t_2(R)]}{\frac{9}{4}(t-\alpha)^2+\Delta} ~,
    \end{equation}
where $t_1(R)$ and $t_2(R)$ are the solutions of \eqref{doble}. Now, let us study this expression for $t-t_1=-\varepsilon$, $\varepsilon>0$ (close but before SCS). By explicitly writing $r(R,t)$ using \eqref{beauty}, we get
\begin{equation}
\partial_R r(R, t_1-\varepsilon)=\frac{3F^{\frac{1}{3}}}{4m(R)}\frac{\varepsilon (\varepsilon+t_2(R)-t_1(R))}{\left[\frac{9}{4}(t_1-\alpha(R)-\varepsilon)^2+\Delta  \right]^{\frac{2}{3}}} ~,
\end{equation}
and, for $\varepsilon <<1$:
\begin{equation}
    \partial_R r(R,t_1-\varepsilon)= \frac{3F^{\frac{1}{3}}}{4m(R)}\frac{\varepsilon(t_2(R)-t_1(R))}{\left[\frac{9}{4}(t_1-\alpha(R))^2+\Delta  \right]^{\frac{2}{3}}}+ O(\varepsilon)~.
\end{equation}
If we define
\begin{equation}
 K(t_1,t_2,R)\equiv \frac{3F}{4m}\frac{(t_2(R)-t_1(R))}{\left[\frac{9}{4}(t_1-\alpha(R))^2+\Delta  \right]^{\frac{2}{3}}}   
\end{equation}
then we can write
\begin{equation}
\partial_R r(R,t)=K(t_1,t_2,R)(t_1-t)+O(t_1-t)~,
\end{equation}
for $t\lesssim t_1$,  with $K(t_1,t_2,R)>0$.

\end{document}